\documentclass{article}
\usepackage{spconf,amsmath,graphicx}
\usepackage{makecell,tabu,booktabs,multirow,multicol,comment,amsmath,amsfonts,cite,xcolor}

\title{Does single-channel speech enhancement improve keyword spotting accuracy? A case study}
\name{Avamarie Brueggeman$^{1}$\sthanks{Work performed at Apple.},
Takuya Higuchi$^{2}$,
    Masood Delfarah$^{2}$,
    Stephen Shum$^{2}$,
    Vineet Garg$^{2}$}
\address{$^{1}$University of Texas at Dallas, USA\\
$^{2}$Apple, USA}

\begin{document}
%
\maketitle
\begin{abstract}
Noise robustness is a key aspect of successful speech applications. Speech enhancement (SE) has been investigated to improve automatic speech recognition accuracy; however, its effectiveness for keyword spotting (KWS) is still under-investigated, especially when only single-channel microphone signals are available. In this paper, we conduct a comprehensive study on single-channel speech enhancement for keyword spotting on the Google Speech Command (GSC) dataset. To investigate robustness to noise, the GSC dataset is augmented with noise signals from the WSJ0 Hipster Ambient Mixtures (WHAM!) noise dataset. Our investigation includes not only applying SE before KWS but also performing joint training of the SE frontend and KWS backend models. Moreover, we explore audio injection, a common approach to reduce distortions by using a weighted average of the enhanced and original signals. Audio injection is then further optimized by using another model that predicts the weight for each utterance. Our investigation reveals that single-channel SE can improve KWS accuracy on noisy speech when the backend model is trained on clean speech. However, despite our extensive exploration, improving the KWS accuracy with single-channel SE is challenging when the backend is trained on noisy speech. We conclude the paper with discussion and future research directions.
\end{abstract}

\section{Introduction}
\label{sec:intro}

Keyword spotting (KWS) is an important technology for speech applications. 
This is particularly the case for virtual assistants on consumer electronics, where a keyword is used to activate the device.
Noise robustness is key for successful KWS since background noise degrades its accuracy.

Speech enhancement (SE) has been investigated to suppress background noise and interfering speakers for improving the noise robustness of downstream tasks \cite{heymann2017beamnet,menne2019investigation,wang2020voicefilter,Sato2022}. However, these investigations focus on automatic speech recognition (ASR) as a downstream task.

In the context of KWS, several papers leverage multichannel SE \cite{haeb2019speech,yu2020end,9054538} and show improvements on KWS accuracies. However, the effect of single-channel SE is still under-investigated for KWS. 
Although recent advances in single-channel SE have shown impressive gains in signal-to-noise ratio (SNR) \cite{6932438,pascual2017segan,luo2019conv,chen2020dual,defossez2020real,subakan2021attention,fu2021metricgan+}
, it is known that improvements in SNR do not always guarantee improvements in downstream tasks, possibly due to artifacts introduced by single-channel non-linear processing. 

In this study, we investigate how single-channel SE works for improving KWS accuracy. We augment the GSC v2 dataset \cite{warden2018speech} with noise signals from the WHAM! noise dataset \cite{wichern2019wham}. Next, we train Conv-TasNet \cite{luo2019conv} on the noisy dataset, perform single-channel SE, and feed enhanced signals to a KWS model, BC-ResNet-8 \cite{kim21l_interspeech}. We also explore joint training of the SE and KWS models to further optimize Conv-TasNet for KWS. In addition, an audio injection approach is investigated to reduce distortions introduced by performing SE. Our investigation reveals that
 SE improves the KWS accuracy when the KWS model is trained on clean data; however, the improvement diminishes when the KWS model is trained on noisy speech. Our results also suggest the need for larger, more realistic datasets for further investigations on noise robust KWS.


\section{PRIOR WORK}
\label{sec:prior}
Noise robustness of KWS models has been investigated in  \cite{9747025,majumdar2020matchboxnet}, where the KWS models are trained and evaluated on noisy data. However, no explicit SE is performed in these investigations. Although explicit SE is performed with multichannel signals for KWS in  \cite{haeb2019speech,yu2020end,9054538}, single-channel SE is still not well-investigated for KWS to the best of our knowledge. Single-channel SE is more challenging than multichannel SE due to the lack of spatial information and artifacts introduced by non-linear processing. In addition, SE for KWS can be more challenging than that for ASR because 1) a keyword is in general shorter than a sentence used for ASR evaluations, and 2) streaming processing is required for popular KWS applications such as wake word detection. Although there were a few papers leveraging single-channel SE for KWS  and showing improvements \cite{yu2018text,gu2019monaural}, there have been advancements since then in both frontend SE and backend KWS models, such as time-domain SE (e.g., \cite{luo2019conv}) and powerful KWS backend models proposed in \cite{kim21l_interspeech}.
In this study, we employ the state-of-the-art KWS acoustic model \cite{kim21l_interspeech} and perform single-channel SE to investigate if it improves KWS accuracy on both clean and noisy conditions.  


\section{Settings}
\label{sec:problem}

\subsection{Datasets}
\label{sec:data}

Our experimental evaluation used the GSC v2 dataset \cite{warden2018speech}, which contains 105,829 one-second utterances of 35 words spoken by over 2,600 speakers. We followed the standard designation of 10 keywords (``yes,'' ``no,'' ``up,'' ``down,'' ``left,'' ``right,'' ``on,'' ``off,'' ``stop,'' ``go''), and the remaining words served as negative examples. There were 12 final classes: ten keywords, one class for unknown words, and one class for silence. We rebalanced utterances of the unknown words and silence classes using the average number of utterances in the remaining classes following common settings in \cite{warden2018speech}. All utterances were single-channel and had a sampling rate of 16 kHz. We followed the standard train/validation/test split from \cite{warden2018speech}.


Due to the limited amount of noise examples, we augmented the GSC v2 dataset with audio from the WHAM! noise dataset \cite{wichern2019wham}, which contains 28,000 files of urban environmental noise from locations such as bars and restaurants (no intelligible speech is present in the dataset).
 All noise samples contained two channels and were sampled at 16 kHz, with an average duration of 10 seconds. We used the first channel to randomly extract one-second segments and added them to the utterances from the GSC v2 dataset. The SNR was randomly chosen from $0$ to $15$ dB. During training, we performed on-the-fly data augmentation to increase the diversity of the training data where the SNR and noise segments were randomly sampled for every training utterance. Only $80\%$ of utterances were augmented on-the-fly following the implementation in \cite{kim21l_interspeech} so that clean speech was also seen during training. We used the official train/validation/test split of \cite{wichern2019wham}. In subsequent results, we refer to the noise augmented data as the noisy GSC v2 dataset. The standard dataset as in \cite{warden2018speech} is referred to as the clean GSC v2 dataset.

\subsection{Speech Enhancement}
\label{sec:SE}
\begin{table}[t]
  \caption{SDR improvements by Conv-TasNet on speech signals from GSC v2 dataset augmented with WHAM! noise dataset.}
  \vspace{3mm}
  \label{tab:se}
   \centering
\begin{tabu}{cc|c}
  \hline
 &  causal & SDRi [dB]\\
  \hline
 \multirow{2}{*}{Conv-TasNet} & yes &5.35\\ 
  & no &5.56\\  \hline
\end{tabu}
\end{table}
In this study, we used Conv-TasNet \cite{luo2019conv} as our frontend SE model. Given the observed noisy signal $\mathbf{x}$, a convolutional encoder first converts the time-domain signal into a 2-D representation. Then a separator predicts a mask that is applied to the representation from the encoder to obtain an enhanced representation. Finally, a convolutional decoder converts the enhanced 2-D representation back to a 1-D time-domain signal.

We used the ESPnet toolkit \cite{lu22c_interspeech} implementation\footnote{https://github.com/espnet/espnet/tree/master/egs2/dns\_icassp22/enh1} of a Conv-TasNet model with decreased kernel and stride lengths of $320$ and $160$, respectively, to keep the same kernel and stride lengths ($20$ and $10$ ms) for 16 kHz signals. The model was trained for 200 epochs on the GSC v2 dataset augmented with WHAM! noise using signal-to-distortion ratio (SDR) as the loss. As in \cite{kim21l_interspeech}, the learning rate was increased linearly from $0$ to $0.1$ for the first five epochs and afterwards decreased to $0$ via cosine annealing. The optimizer was stochastic gradient descent with momentum set to $0.9$ and a weight decay of $0.001$. The SDR improvements on the noisy GSC v2 dataset are shown in Table~\ref{tab:se} for our causal and non-causal Conv-TasNet models. In the remainder of this paper, we utilize the causal Conv-TasNet model due to considerations for streaming applications. 

\subsection{Keyword Spotting (KWS)}
\label{sec:KWS}
\begin{table}[t]
  \caption{Baseline KWS accuracies on clean and noisy GSC v2 test sets without speech enhancement.}
  \vspace{3mm}
  \label{tab:baseline}
   \centering
\begin{tabu}{cc|ccc}
  \hline
    & &  \multicolumn{3}{c}{test acc. [\%]} \\
 model ID &  train data & clean& noisy & avg.\\
  \hline
 M1 & clean &\textbf{98.7}&94.3 &96.5\\ 
 M2 & noisy &98.4&\textbf{96.0}&\textbf{97.2}\\  \hline
\end{tabu}
\vspace{-0.25cm}
\end{table}
For the KWS backend, we trained a BC-ResNet model \cite{kim21l_interspeech} following the official implementation\footnote{https://github.com/Qualcomm-AI-research/bcresnet}.
 We used the largest model available, BC-ResNet-8, to take full advantage of the powerful backend model. For more details on the backend model, refer to \cite{kim21l_interspeech}. As shown in Table~\ref{tab:baseline}, the model M1 obtains state-of-the-art performance on the clean (unaltered) GSC v2 test set with $98.7$\% accuracy.
 To create an additional baseline, we trained another BC-ResNet-8 model on the GSC v2 dataset augmented with WHAM! noise to produce model M2.
 While M1 achieved the state-of-the-art performance on the clean test set, the accuracy was degraded on the noisy set. By training on noisy data, M2 achieved better accuracy on the noisy test set, while a marginal regression was observed on the clean set.


\subsection{Joint Training}
\label{sec:JT}
An SE model can be sub-optimal for KWS when trained using a signal-level loss function such as SDR. Joint training of both the SE and KWS models can naturally be performed via back-propagation. We explored the use of KWS-only loss or combined SDR and KWS loss functions. In the former case, the frontend and backend models were trained for 100 epochs using the cross-entropy (CE) loss for keywords.
For the latter case, the CE loss was combined with the SDR loss using a scale factor $\beta$ as 
\begin{align}
L = L_{CE} + \beta * L_{SDR}, \label{eq:combloss}
\end{align}
where $\beta$ was set to $0.01$, which was tuned on the noisy validation set. A learning rate was tuned separately for each model on the noisy validation set and reported in Section \ref{sec:exp}.


\subsection{Audio Injection}
\label{sec:SW}
Audio injection is a common technique to reduce distortions introduced by signal processing. Let $\mathbf{x}'$ and $\mathbf{x}$ denote the enhanced and original noisy signal, respectively. The weighted sum is computed as
\begin{align}
\mathbf{x}'' = \alpha \mathbf{x}' + (1-\alpha) \mathbf{x}, \label{eq:ai}
\end{align}
where $\alpha$ is a global hyper-parameter ranging from 0 to 1 that is applied to all utterances. Then, $\mathbf{x}''$ is fed into the KWS backend model.

Optionally, $\alpha$ can be predicted for each utterance via neural networks as proposed in \cite{wang2020voicefilter,Sato2022}. Following the settings in \cite{Sato2022}, a soft-switching model, another neural network, takes $\mathbf{x}$ and $\mathbf{x}'$ and predicts the optimal $\alpha$ for each utterance independently. We used 256-dimensional log mel-filterbank features with three bidirectional Long Short-Term Memory (BLSTM) layers, each of which has 128 hidden units, followed by attention pooling, two fully-connected layers with 128 hidden units, and a soft-max layer. The model was trained with the KWS CE loss for 20 epochs with a learning rate of $0.01$.





\section{Results}
\label{sec:exp}






\begin{table*}[t]
  \caption{SDRi on noisy GSC test set and KWS accuracies on clean and noisy GSC test sets with SE.}
  \label{tab:joint}
   \centering
   \scalebox{0.8}{
\begin{tabu}{cccc|cccc}
  \hline
    & & & & &\multicolumn{3}{c}{test acc. [$\%$]} \\
 backend model ID & learning rate& freeze frontend & freeze backend & SDRi [dB] & clean& noisy & avg.\\
  \hline
 M1 && yes & yes & 5.35 &\textbf{98.1}&92.1 & 95.1\\ 
 M2 &&yes & yes & 5.35 &\textbf{98.1}&93.0& 95.6\\
 \hline\hline
  M1 &1e-4& no & yes & -3.66 & 97.5&94.0&95.8\\ 
  M1 &1e-3& yes & no & 5.35 &98.0&95.4&\textbf{96.7}\\ 
  M1 &1e-4& no & no & -1.30 &97.5&95.2&96.4\\ 
 \hline\hline
  M2 &1e-5& no & yes & 1.36 & 97.6&94.9&96.3\\ 
  M2 &1e-3& yes & no & 5.35 &98.0&95.1&96.6\\ 
  M2 &1e-5& no & no & 2.07 &97.8&95.5&96.7\\ 
 M2&1e-5& \multicolumn{2}{c|}{+combined loss ($\beta$=0.01)} & 4.96 &97.8&\textbf{95.8}&\textbf{96.8}\\ 
 \hline
\end{tabu}
}
\vspace{-3mm}
\end{table*}

\begin{figure}[t]
\begin{center}
\includegraphics[width=7cm]{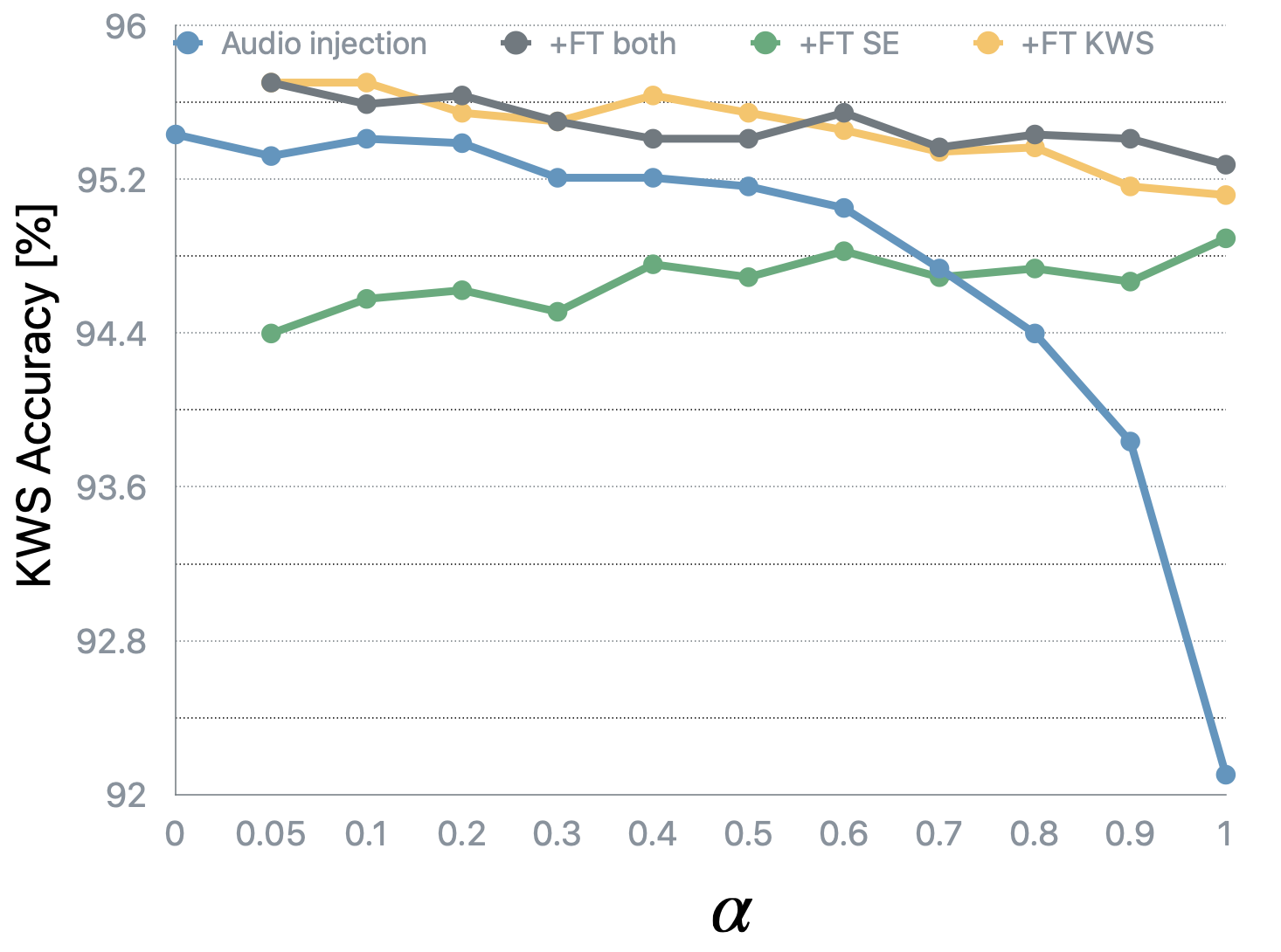}
%
\vspace{-0.5cm}
\caption{KWS accuracies on noisy validation set.}
\label{fig:valid}
\vspace{-0.75cm}
\end{center}
\end{figure}

\begin{figure}[t]
\begin{center}
\includegraphics[width=7cm]{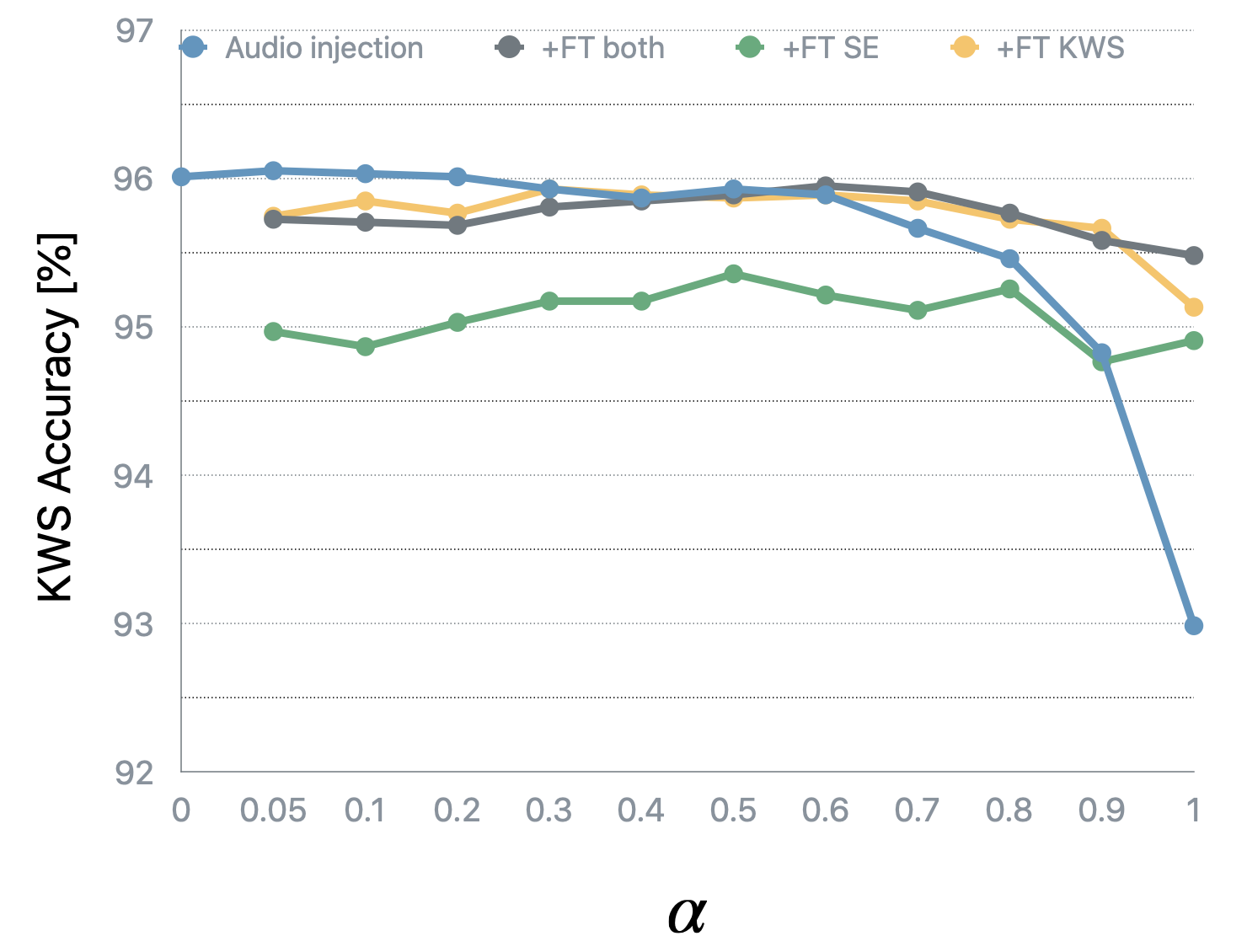}
%
\vspace{-0.5cm}
\caption{KWS accuracies on noisy test set.}
\label{fig:test}
\vspace{-0.75cm}
\end{center}
\end{figure}

\begin{table*}[t]
  \caption{SDRi on noisy GSC test set and KWS accuracies on clean and noisy GSC test sets with SE and audio injection.}
  \label{tab:injection}
   \centering
   \scalebox{0.8}{
\begin{tabu}{ccccc|cccc}
  \hline
    & & & & & &\multicolumn{3}{c}{test acc. [$\%$]} \\
 backend model ID & learning rate& freeze frontend & freeze backend & $\alpha$ &SDRi [dB] & clean& noisy & avg.\\
  \hline
 M2 && yes & yes &$0.1$& 0.59 &\textbf{98.6}&\textbf{96.0} & \textbf{97.3}\\   \hline\hline
  M2 &1e-5& no & yes &$0.6$& 0.67 &97.4&95.2 & 96.3\\ 
  M2 &1e-5& yes & no &$0.1$& 0.59&98.5&95.8 & 97.2\\ 
  M2 &1e-5& no & no &$0.05$& 0.28 &98.4&95.7 & 97.1\\ 
 \hline\hline
 \rowfont{\color{gray}}
 M2 &1e-2& yes & yes &predicted \cite{Sato2022}& 9.19 &97.8&95.0 & 96.4\\
 \hline
\end{tabu}
}
\vspace{-3mm}
\end{table*}


Table \ref{tab:joint} shows KWS accuracies with SE. By performing SE, regressions were observed with both M1 and M2 on clean and noisy sets. However, by jointly training the backend model, the accuracy with M1 on the noisy set was improved from $94.3\%$ without SE (baseline) to $95.4\%$ with SE. This result shows the effectiveness of SE when the backend model was trained on the clean data. However, with M2, we saw no improvement compared to the baseline  ($96.0\%$) reported in Table \ref{tab:baseline} even after joint training.
The combined loss slightly improved the accuracy on the noisy set; however, the baseline without SE still outperformed the systems with SE. 

Figures \ref{fig:valid} and \ref{fig:test} show KWS performance on the noisy validation and test sets, respectively, with the audio injection approach and joint training. ``+FT both," ``+FT SE," and ``+FT KWS" refer to audio injection with fine-tuning both frontend and backend models, only the frontend model, and only the backend model, respectively. $\alpha=0$ indicates the result without SE, and $\alpha=1$ indicates the result from simply using the output from SE (no audio injection). Without fine-tuning, the KWS accuracy improved as $\alpha$ was lowered from 1; however, no meaningful improvement was observed compared to the baseline result with $\alpha=0$. On the validation set, fine-tuning the models yielded better accuracies, especially with high values for $\alpha$, and outperformed the baseline accuracy ($\alpha=0$). However, no meaningful improvement can still be seen on the test set. Table \ref{tab:injection} shows results with the best $\alpha$ tuned based on the validation accuracy. The audio injection approach without fine-tuning matched the baseline in terms of the accuracy on noisy test set, and slightly outperformed for the clean test set. The soft-switching model proposed in \cite{Sato2022} did not improve the accuracy in our experiments even though the model was trained to predict optimal $\alpha$ for a better KWS accuracy.

\section{Discussion}
\label{sec:disc}
Despite our extensive exploration, speech enhancement did not improve the KWS accuracy when the backend model was trained on noisy data. In this section, we discuss the potential root cause and future research directions.

\subsection{Limited Context for SE}
\label{sec:limitedcontext}
A dilated convolution in Conv-TasNet enables a long receptive field \cite{luo2019conv}, which in general is useful because the model can learn speech and noise characteristics from a long context of audio. However, each utterance is segmented in the GSC dataset and its duration is only one second.
Because of this, the speech enhancement performance may be limited when the model is trained and evaluated on utterances from the GSC dataset. The comparison of causal and non-causal models (See Table \ref{tab:se}) also suggests that each utterance would contain limited context for SE as the improvement with the non-causal model was limited compared to the causal model.
 It should also be noted that reducing the receptive field with a smaller kernel size of the encoder did not yield improvement in the SDR in our preliminary experiment.
Further investigation may be required with more practical training and evaluation data, e.g., streaming audio with a long context, to take full advantage of the SE model.

\subsection{Limited Amount of Data for Joint Training}
\label{sec:limitedamount}
Joint training improved KWS accuracies on the validation set as shown in Fig.~\ref{fig:valid}; however, the jointly trained models did not improve the accuracies on the test set (Fig. \ref{fig:test}). This implies that the models are not well-generalized for the held-out test set, possibly due to the limited amount of data in the GSC dataset used for joint training. A larger KWS dataset may be required for further investigations on joint training.  

\subsection{Real Versus Simulated Noisy Data}
\label{sec:realrec}
In our experiments, we used simulated noisy data instead of real recordings for backend training. In practice, real noisy recordings can be used for backend model training as long as the recordings are annotated for keywords. This allows us to train the backend model on in-domain real noisy data, which potentially further improves the KWS accuracy compared to the backend model trained on simulated noisy data as done in the experiments. On the other hand, pairs of clean and noisy data are required to train the frontend model, which prevents us from using in-domain real recordings. Thus, there could be a potential mismatch between training and test data for SE. Several approaches have been proposed and investigated to overcome this limitation \cite{wisdom2020unsupervised,tzinis2022remixit,leglaive2023chime}; however, further investigations are required to validate their effectiveness on practical KWS where real noisy recordings are available for training. 

\subsection{Speech Versus Non-speech Noise}
\label{sec:nonspeech}
Speech noise can also be a future research direction for KWS. Our investigation has been done only with non-speech noise while the effectiveness of SE has been shown for ASR with speech noise in \cite{wang2020voicefilter,Sato2022}. In contrast to ASR, KWS models might still be able to handle speech noise since non-keyword speech noise could simply be ignored. 

\vspace{-0.1cm}
\section{Conclusions}
\label{sec:conc}
\vspace{-0.2cm}

In this study, we have explored single-channel SE for improving KWS accuracy in noisy conditions. Although SE improved KWS accuracy when the backend KWS model was trained on clean speech, no meaningful improvement was observed when the backend model was trained on noisy speech. Joint training of SE and KWS models and/or the audio injection approach improved KWS accuracy compared to applying the pre-trained SE model. However, the KWS model trained on noisy speech still achieved the best performance on the noisy test set when SE was not performed. The soft-switching approach also did not improve the accuracy in our study. It is suggested that larger and more realistic datasets may be required for further investigation on SE for KWS. 

%

\vspace{-0.1cm}
\section{Acknowledgement}
\label{sec:ack}
\vspace{-0.2cm}
We thank Mehrez Souden for his feedback on the paper and the helpful discussions.

\vfill\pagebreak


\bibliographystyle{IEEEbib_short}
\bibliography{refs}

\begin{thebibliography}{10}

\bibitem{heymann2017beamnet}
Jahn Heymann, et~al.,
\newblock ``Beamnet: End-to-end training of a beamformer-supported
  multi-channel asr system,''
\newblock in {\em 2017 IEEE International Conference on Acoustics, Speech and
  Signal Processing (ICASSP)}. IEEE, 2017, pp. 5325--5329.

\bibitem{menne2019investigation}
Tobias Menne, et~al.,
\newblock ``Investigation into joint optimization of single channel speech
  enhancement and acoustic modeling for robust asr,''
\newblock in {\em ICASSP 2019-2019 IEEE International Conference on Acoustics,
  Speech and Signal Processing (ICASSP)}. IEEE, 2019, pp. 6660--6664.

\bibitem{wang2020voicefilter}
Quan Wang, et~al.,
\newblock ``Voicefilter-lite: Streaming targeted voice separation for on-device
  speech recognition,''
\newblock {\em arXiv preprint arXiv:2009.04323}, 2020.

\bibitem{Sato2022}
Hiroshi Sato, et~al.,
\newblock ``Learning to enhance or not: Neural network-based switching of
  enhanced and observed signals for overlapping speech recognition,''
\newblock in {\em ICASSP 2022 - 2022 IEEE International Conference on
  Acoustics, Speech and Signal Processing (ICASSP)}, 2022, pp. 6287--6291.

\bibitem{haeb2019speech}
Reinhold Haeb-Umbach, et~al.,
\newblock ``Speech processing for digital home assistants: Combining signal
  processing with deep-learning techniques,''
\newblock {\em IEEE Signal processing magazine}, vol. 36, no. 6, pp. 111--124,
  2019.

\bibitem{yu2020end}
Meng Yu, et~al.,
\newblock ``End-to-end multi-look keyword spotting,''
\newblock {\em arXiv preprint arXiv:2005.10386}, 2020.

\bibitem{9054538}
Xuan Ji, et~al.,
\newblock ``Integration of multi-look beamformers for multi-channel keyword
  spotting,''
\newblock in {\em ICASSP 2020 - 2020 IEEE International Conference on
  Acoustics, Speech and Signal Processing (ICASSP)}, 2020, pp. 7464--7468.

\bibitem{6932438}
Yong Xu, et~al.,
\newblock ``A regression approach to speech enhancement based on deep neural
  networks,''
\newblock {\em IEEE/ACM Transactions on Audio, Speech, and Language
  Processing}, vol. 23, no. 1, pp. 7--19, 2015.

\bibitem{pascual2017segan}
Santiago Pascual, et~al.,
\newblock ``Segan: Speech enhancement generative adversarial network,''
\newblock {\em arXiv preprint arXiv:1703.09452}, 2017.

\bibitem{luo2019conv}
Yi~Luo and Nima Mesgarani,
\newblock ``Conv-tasnet: Surpassing ideal time--frequency magnitude masking for
  speech separation,''
\newblock {\em IEEE/ACM Transactions on Audio, Speech, and Language
  Processing}, vol. 27, no. 8, pp. 1256--1266, 2019.

\bibitem{chen2020dual}
Jingjing Chen, et~al.,
\newblock ``Dual-path transformer network: Direct context-aware modeling for
  end-to-end monaural speech separation,''
\newblock {\em arXiv preprint arXiv:2007.13975}, 2020.

\bibitem{defossez2020real}
Alexandre Defossez, et~al.,
\newblock ``Real time speech enhancement in the waveform domain,''
\newblock {\em arXiv preprint arXiv:2006.12847}, 2020.

\bibitem{subakan2021attention}
Cem Subakan, et~al.,
\newblock ``Attention is all you need in speech separation,''
\newblock in {\em ICASSP 2021-2021 IEEE International Conference on Acoustics,
  Speech and Signal Processing (ICASSP)}. IEEE, 2021, pp. 21--25.

\bibitem{fu2021metricgan+}
Szu-Wei Fu, et~al.,
\newblock ``Metricgan+: An improved version of metricgan for speech
  enhancement,''
\newblock {\em arXiv preprint arXiv:2104.03538}, 2021.

\bibitem{warden2018speech}
Pete Warden,
\newblock ``Speech commands: A dataset for limited-vocabulary speech
  recognition,''
\newblock {\em arXiv preprint arXiv:1804.03209}, 2018.

\bibitem{wichern2019wham}
Gordon Wichern, et~al.,
\newblock ``Wham!: Extending speech separation to noisy environments,''
\newblock {\em arXiv preprint arXiv:1907.01160}, 2019.

\bibitem{kim21l_interspeech}
Byeonggeun Kim, et~al.,
\newblock ``{Broadcasted Residual Learning for Efficient Keyword Spotting},''
\newblock in {\em Proc. Interspeech 2021}, 2021, pp. 4538--4542.

\bibitem{9747025}
Dianwen Ng, et~al.,
\newblock ``Convmixer: Feature interactive convolution with curriculum learning
  for small footprint and noisy far-field keyword spotting,''
\newblock in {\em ICASSP 2022 - 2022 IEEE International Conference on
  Acoustics, Speech and Signal Processing (ICASSP)}, 2022, pp. 3603--3607.

\bibitem{majumdar2020matchboxnet}
Somshubra Majumdar and Boris Ginsburg,
\newblock ``Matchboxnet: 1d time-channel separable convolutional neural network
  architecture for speech commands recognition,''
\newblock {\em arXiv preprint arXiv:2004.08531}, 2020.

\bibitem{yu2018text}
Meng Yu, et~al.,
\newblock ``Text-dependent speech enhancement for small-footprint robust
  keyword detection.,''
\newblock in {\em Interspeech}, 2018, pp. 2613--2617.

\bibitem{gu2019monaural}
Yue Gu, et~al.,
\newblock ``A monaural speech enhancement method for robust small-footprint
  keyword spotting,''
\newblock {\em arXiv preprint arXiv:1906.08415}, 2019.

\bibitem{lu22c_interspeech}
Yen-Ju Lu, et~al.,
\newblock ``{ESPnet-SE++: Speech Enhancement for Robust Speech Recognition,
  Translation, and Understanding},''
\newblock in {\em Proc. Interspeech 2022}, 2022, pp. 5458--5462.

\bibitem{wisdom2020unsupervised}
Scott Wisdom, et~al.,
\newblock ``Unsupervised sound separation using mixture invariant training,''
\newblock {\em Advances in Neural Information Processing Systems}, vol. 33, pp.
  3846--3857, 2020.

\bibitem{tzinis2022remixit}
Efthymios Tzinis, et~al.,
\newblock ``Remixit: Continual self-training of speech enhancement models via
  bootstrapped remixing,''
\newblock {\em IEEE Journal of Selected Topics in Signal Processing}, vol. 16,
  no. 6, pp. 1329--1341, 2022.

\bibitem{leglaive2023chime}
Simon Leglaive, et~al.,
\newblock ``The chime-7 udase task: Unsupervised domain adaptation for
  conversational speech enhancement,''
\newblock {\em arXiv preprint arXiv:2307.03533}, 2023.

\end{thebibliography}

\end{document}